# A-site Cation Influence on the Conduction Band of Lead Bromide Perovskites

(dated Sept. 17, 2021)

Gabriel J. Man*[1], Chinnathambi Kamal[2,3], Aleksandr Kalinko[4], Dibya Phuyal[5], Joydev Acharya[6], Soham Mukherjee[1], Pabitra K. Nayak[6], Håkan Rensmo[1], Michael Odelius[2] and Sergei M. Butorin*[1]

**Author addresses:**

1. Condensed Matter Physics of Energy Materials, Division of X-ray Photon Science, Department of Physics and Astronomy, Uppsala University, Box 516, Uppsala 75121, Sweden
2. Department of Physics, Stockholm University, AlbaNova University Center, Stockholm 10691, Sweden
3. Theory and Simulations Laboratory, HRDS, Raja Ramanna Centre for Advanced Technology, Indore 452013, India
4. Photon Science DESY, Notkestraße 85, Hamburg 22607, Germany
5. Division of Material and Nano Physics, Department of Applied Physics, KTH Royal Institute of Technology, Stockholm 10691, Sweden
6. TIFR Centre for Interdisciplinary Sciences, Tata Institute of Fundamental Research, 36/P, Gopanpally Village, Serilingampally Mandal, Hyderabad 500046, India

* To whom correspondence should be addressed.  gman@alumni.princeton.edu ,
sergei.butorin@physics.uu.se

## Abstract

Hot carrier solar cells hold promise for exceeding the Shockley-Queisser limit.  Slow hot carrier cooling is one of the most intriguing properties of lead halide perovskites and distinguishes this class of materials from competing materials used in solar cells.  Here we use the element selectivity of high-resolution X-ray spectroscopy to uncover a previously hidden feature in the conduction band states, the σ-π energy separation, and find that it is strongly influenced by the strength of electronic coupling between the A-cation and bromide-lead sublattice.  Our finding provides an alternative mechanism to the commonly discussed polaronic screening and hot phonon bottleneck carrier cooling mechanisms.  Our work emphasizes the optoelectronic role of the A-cation, provides a comprehensive view of A-cation effects in the electronic and crystal structures, and outlines a broadly applicable spectroscopic approach for assessing the impact of chemical alterations of the A-cation on halide and potentially non-halide perovskite electronic structure.

**Introduction**

Lead halide perovskites (HaP) of the form $APbX_3$ have attracted renewed research interest for over a decade, motivated by initially dramatic gains in HaP solar cell efficiencies and now other optoelectronic applications[1–5]. The prototypical A-site cations (A-cations) are organic (methylammonium or $MA^+$, formamidinium or $FA^+$) or inorganic ($Cs^+$), the B-site cation is lead(II) and the X-site anion is iodide/bromide/chloride. At present, in spite of the substantial growth of many new subclasses of HaP-related materials and their applications, many fundamental questions related to the prototypical HaPs remain unanswered, despite a history of basic research dating back to as early as the late 1970's[6,7]. One such question concerns the optoelectronic function of the A-cation[8]. Studies based on a range of complementary approaches: time-resolved photoluminescence (PL), combined electron spectroscopy and partial density of states (PDOS) calculations, mechanical nanoindentation and solar cell characterization have yielded evidence for and against the existence of A-cation optoelectronic functionality[9–13].

To unravel this conundrum, we utilize element- and orbital-selective core level spectroscopy which includes X-ray absorption spectroscopy (XAS) for probing conduction band states and resonant X-ray emission spectroscopy (RXES) for probing valence band states[14]. The energy resolution of conventional XAS is intrinsically limited by core-hole lifetime broadening, which may obscure conduction band features probed in the spectra[15]. We circumvent this limitation by using High Energy Resolution Fluorescence Detected XAS (HERFD-XAS) which yields reduced broadenings of 2.2 eV (vs. 2.5 eV) and 2.5 eV (vs. 6.1 eV) for the bromine $K$-edge (*1s→p* transition) and lead $L_3$-edge (*2p_{3/2}→s,d* transition) absorption spectra, respectively[15,16]. Only two reports of HERFD-XAS, applied to the study of HaP thin films, exist, with just one report examining the prototypical compounds methylammonium lead tri-iodide/bromide (MAPI/B) [17,18]. In this work, we report a joint experimental and computational investigation of three prototypical lead bromide perovskites (APB): MAPB, formamidinium lead tribromide (FAPB), cesium lead tribromide (CsPB), facilitated by single crystals which are durable under X-ray irradiation. Density functional theory (DFT) was employed for *ab initio* molecular dynamics simulations (AIMD) combined with ground-state PDOS and bromine *K*-edge XAS calculations. The use of HERFD-XAS enables us to differentiate, in terms of electronic structure, between compounds with the same lead/bromide formal oxidation states and ultimately attribute measured differences to A-cation type. We find the strength of electronic coupling between the A-cation and bromide-lead sublattice increases in this order: $Cs^+ \rightarrow MA^+ \rightarrow FA^+$, and influences the energetic width of the conduction band. We discuss the connection between conduction band width and slow cooling of hot electrons. We observe that higher coupling strength correlates with higher Br-Pb bond ionicity and a decreased energy offset between a given energy level and Br *1s*, and link this finding to energy level matching at optoelectronic device interfaces.

**Results**

We first focus on the unoccupied conduction band states as the occupied states of HaPs have been investigated more frequently, possibly due to the higher availability of commercial photoelectron spectroscopy (PES) instrumentation. We experimentally profile the unoccupied states in an element- and orbital-resolved manner via HERFD-XAS spectra derived from RXES maps. The maps are generated with a synchrotron beamline and a crystal-based X-ray spectrometer [19,20]. A representative Br *K* RXES map, recorded from single crystal MAPB, is displayed in Fig. 1a and features resonant and off-resonant X-ray emission from three *p*→*s* transitions: $K\beta_{1,3}$ ($3p_{3/2}/3p_{1/2} \rightarrow 1s$) and $K\beta_2$ ($4p \rightarrow 1s$). An energy level schematic is shown in Fig. 1b and depicts the one-electron transitions related to X-ray absorption into conduction band states and emission from valence band and shallow core level states. The final state of valence photoemission is shown for reference. A representative Pb $L_3$ RXES map is displayed in Supplementary Figure (SF) 1 and the analysis is described in Supplementary Note (SN) 1. Vertical dashed lines in all RXES maps represent constant-emission-energy map cuts through the maximum of the RXES intensity that yield the HERFD-XAS spectra. For material systems with relatively delocalized states, such a RXES map cut has been shown to be a good high-resolution approximation to the conventional XAS spectrum[21]. The conduction band states of interest are profiled in the near-edge features, which we enclose with a box marked region of interest (ROI).

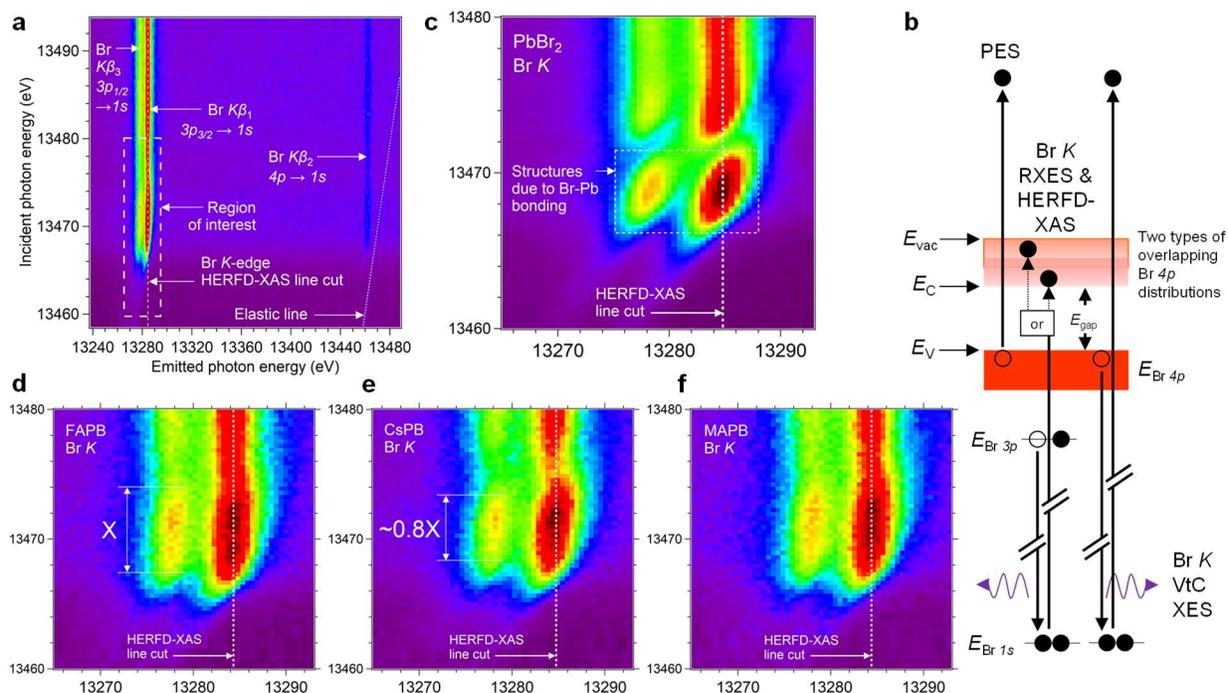

**Figure 1 | Bromine $K$-edge resonant XES full map and region of interest.** The HERFD-XAS line cuts are visualized as dashed vertical white lines in all maps. **a,** Representative Br $K$ full map, recorded from single crystal MAPB. The map shows two core-to-core transitions ($K\beta_{1,3}$) and the valence-to-core transition (VtC, $K\beta_2$), on- and off-resonance. The elastic line is marked with a guide for the eye. **b,** Energy level diagram depicting the core hole decay processes measured in the RXES map, along with the valence photoemission final state for comparison. The valence and conduction band edges are labelled $E_V$ and $E_C$, respectively, $E_{gap}$ denotes the bandgap and $E_{vac}$ denotes the ionization threshold or vacuum level. **c-f,** Regions of interest of PbBr$_2$, FAPB, CsPB and MAPB. The x- and y-axis labels for the regions of interest are the same as the full map. Relative main-edge widths in d,e are marked with "X" and "0.8X".

The Br $K$ ROIs for PbBr$_2$, FAPB, CsPB and MAPB are displayed in Fig. 1c-f.  The A-cation has an observable effect on the near-edge features.  The ROI for PbBr$_2$ shows the first HERFD-XAS feature from ~13466 to ~13471 eV in incident energy.  The ROIs for the APB compounds show HERFD-XAS features in the same energetic region as PbBr$_2$ and up to ~13475 eV, suggesting the higher energy part of the HERFD-XAS feature is affected by bromide interaction with the A-cation.  We observe a difference in the energetic width of the HERFD-XAS feature between FAPB and CsPB, where the width of the CsPB feature is ~80% of the width of the FAPB feature.  Despite the influence of the Br $1s$ core hole present in the final state of the spectroscopic process, the Br $K$ spectrum can be conceptually viewed as a reflection of the unoccupied Br $p$-state distribution.  The HERFD-XAS feature at lowest photon energy is related to the part of the unoccupied states close to the conduction band minimum (CBM), relevant for optoelectronic device operation.  Hence we deduce that the A-cation influences the (i) width and (ii) bonding character of the higher energy part of the conduction band in APB compounds.

We plot Br $K$ HERFD-XAS cuts of PbBr$_2$, FAPB, MAPB and CsPB as one-dimensional spectra in Fig. 2a for quantitative analysis.  Conventional XAS spectra recorded in total fluorescence yield (TFY) mode are shown for comparison.  The resolution improvement of Br $K$ HERFD-XAS (2.2 eV) versus TFY-XAS (2.5 eV) is small but is still important for resolving spectral differences between APB compounds in the energy region between ~13466 to ~13475 eV.  For example, the notch in the spectrum of FAPB (~13470.3 eV, Fig. 2b), resolved with HERFD-XAS but not with TFY-XAS, indicates that the main absorption feature (main-edge) is formed of two components.  All four Br $K$ HERFD-XAS spectra show a feature at ~13469 eV.  Additionally, the spectra for the APB compounds show features around 13471.5 eV which are affected by A-cation replacement, as noted above.  We further confirm this deduction using calculations.  The Pb $L_3$ ROIs and HERFD-XAS spectra are shown in SF 1 and the analysis is described in SN 1.  We use Pb $L_3$ as a direct spectroscopic probe of changes in the electronic structure caused by crystal structure changes and the resolution enhancement yielded by HERFD-XAS (2.5 vs. 6.1 eV) is crucial.

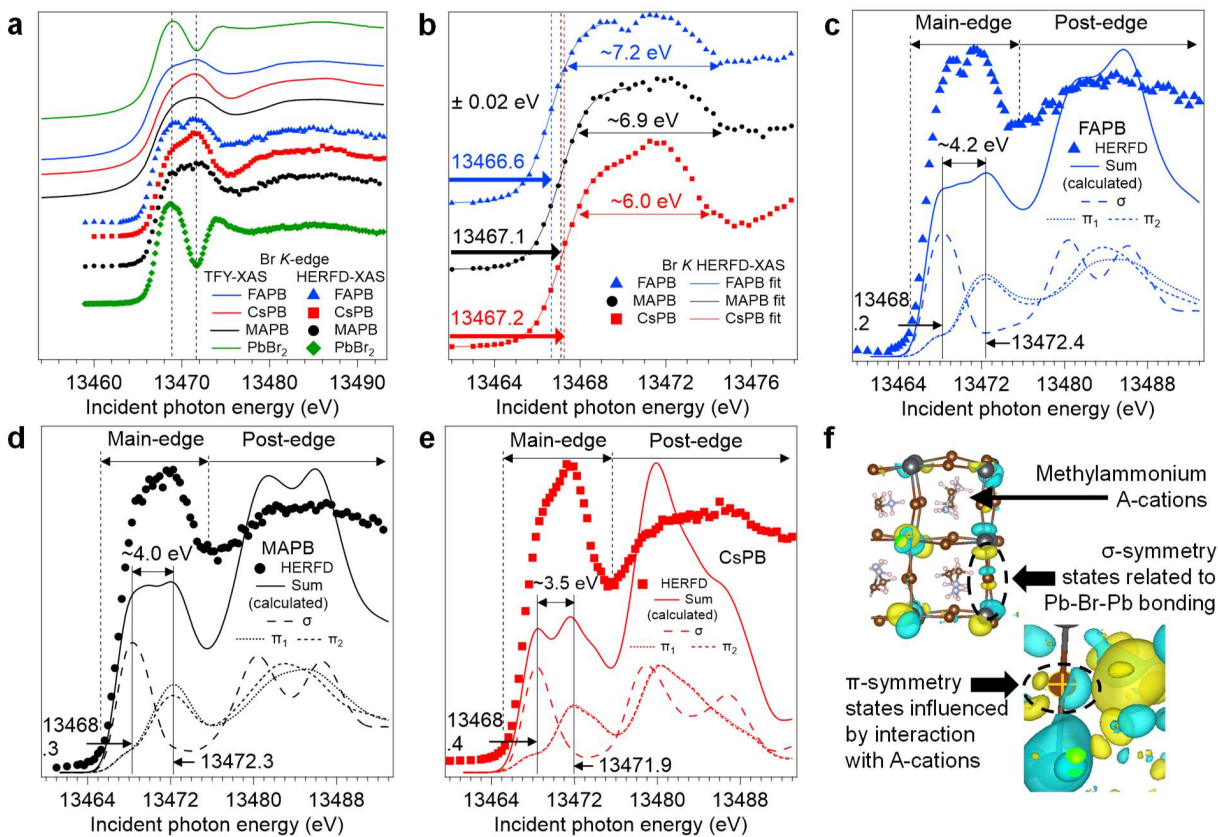

**Figure 2 | Experimental and calculated Br *K*-edge XAS spectra of FAPB, MAPB and CsPB. a**, Comparison of the TFY-XAS (lines) and HERFD-XAS (symbols) spectra of PbBr₂ (green), FAPB (blue), MAPB (black) and CsPB (red). The vertical dashed lines are guides to the eye for features of interest. **b**, Comparison of the HERFD-XAS spectra of the APB compounds. The absorption onsets are quantified with sigmoid fits and the main-edge widths are estimated (see text). **c-e,** Experimental versus calculated Br *K* XAS spectra for FAPB, MAPB and CsPB. The total calculated spectrum is shown along with its constituent distributions. The sigma- and pi-symmetry distributions of states are denoted as σ, π₁ and π₂. **f,** Crystal structure of MAPB, shown with the molecular orbitals associated with σ- and π-symmetry states probed with XAS.

We sigmoid-fit the Br *K* HERFD-XAS spectra of FAPB, MAPB and CsPB to extract numerical parameters (Fig. 2b). Given the signal-to-noise level of our measurements and nearly identical absorption onsets for MAPB and CsPB, we find the sigmoid fit better suited for differentiating the main-edge positions as compared to the first derivative (SF 2). We find an absorption onset trend, based on edge positions associated with a 0.5 step height, which increases in this order: FAPB (13466.6 eV), MAPB (13467.1 eV), CsPB (13467.2 eV). The sigmoid fit uncertainty is 20 meV. The A-cation influences the absorption onsets, suggesting the A-cation affects device-relevant conduction band energy level positions referenced to Br *1s* in the ground-state. The Br *K* spectra of FAPB and MAPB show a post-edge region at ~13476 eV and higher, at a sigmoid step height of 0.75. Using this step height to systematically compare main-edge widths, we estimate widths of ~7.2 eV (FAPB), ~6.9 eV (MAPB) and ~6.0 eV (CsPB).

To understand the origin of the spectral differences and link the crystal and electronic structures, we carry out periodic DFT calculations and AIMD simulations and sample calculated Br *K* XAS spectra using the transition potential DFT method with the half-core hole approximation[22]. In Fig. 2c-e, we compare the calculated Br *K* XAS spectra, averaged over different configurations and bromide sites, against experiment. Through an analysis of the Cartesian spectral components in the crystal frame, we approximately distinguish the transitions to states of σ-symmetry in the Pb-Br-Pb direction from those of π-symmetry (as the Pb-Br-Pb is not perfectly linear) and average them separately, taking into account the crystal orientations of the Br-Pb bonds. Representative molecular orbitals for σ- and π-symmetry in MAPB are visualized in Fig. 2f. From the XAS component spectra, we observe that the rising part of the main-edge can be assigned to Br *4p* states with σ-symmetry while features in the higher energy region of the main-edge are contributed by Br *4p* states with π-symmetry, in all cases. We also notice that due to the spatial extent of the states of π-symmetry, they are more strongly influenced by A-cation interactions. The lower energy position of the σ-symmetry states suggests the states in the vicinity of the CBM in the ground-state are primarily of Br-Pb σ anti-bonding character.  Additional details on the comparison between experimental and calculated XAS are presented in SN 2.

We observe a trend in the energetic separation (σ-π offset) between the maxima of the σ- and π-symmetry distributions in the main-edge: ~3.5 eV (CsPB) → ~4.0 eV (MAPB) → ~4.2 eV (FAPB).  We compare relative energy separation from calculation versus relative main-edge width from experiment. The CsPB : FAPB ratios are ~0.83 (calculation) and ~0.83 (experiment) and the MAPB : FAPB ratios are ~0.95 (calculation) and ~0.96 (experiment).  We find a positive and potentially linear correlation between the relative measures of main-edge width and conclude that the A-cation influences the Br $K$ main-edge width of the APB compounds via the σ-π offset.  The use of HERFD-XAS has revealed A-cation-induced differences in APB electronic structure, and the close agreement between experimental and calculated XAS trends offers an opportunity to computationally elucidate the σ-π offset mechanism.

We study the Br-(A-cation) interaction using calculated Br PDOS from ground-state Kohn-Sham orbitals.  For FAPB and MAPB, we choose two geometries around Br atoms, representing strong (H-Br bond distances of ~2.37 (FAPB), 2.47 Å (MAPB)) and weak (H-Br bond distance > 3.0 Å) hydrogen bonds. We calculate the approximate σ- and π-symmetry contributions to the Br PDOS along Pb-Br-Pb in the crystal frame (remembering there is a bend in Pb-Br-Pb) and connect them with XAS σ- and π-symmetry contributions.  The Br $4s,p$ PDOS for strong hydrogen bonds is presented in Fig. 3a,b; the PDOS for weak bonds is presented in SF 3.  We observe that the occupied Br $4p$ π from -4 to 0.7 eV and Br $4s$ from -15 to -11.5 eV for CsPB, and MAPB/FAPB with short/long H-Br bonds are similar.  The $4p$ σ from -4.5 to 0.5 eV shows some dependence on the A-cation type.  On the other hand, the unoccupied Br $4p$ PDOS exhibits striking differences.  The bands of π-symmetry are shifted and change shapes for FAPB or MAPB with respect to CsPB (between ~4 to ~16 eV); this clearly shows the influence of hydrogen bonding.  We observe that the difference for FAPB versus CsPB is larger than for MAPB versus CsPB, in agreement with the increased hydrogen bond strength (H-Br bond distance is shorter for FAPB).  We quantify the ground-state σ-π offset with peak and center-of-gravity fits (Supplementary Table (ST) 1) and find, irrespective of the H-Br bond distance, that the σ-π offset increases going from CsPB → MAPB → FAPB. This is in qualitative agreement with the σ-π offset trend in the XAS spectra.

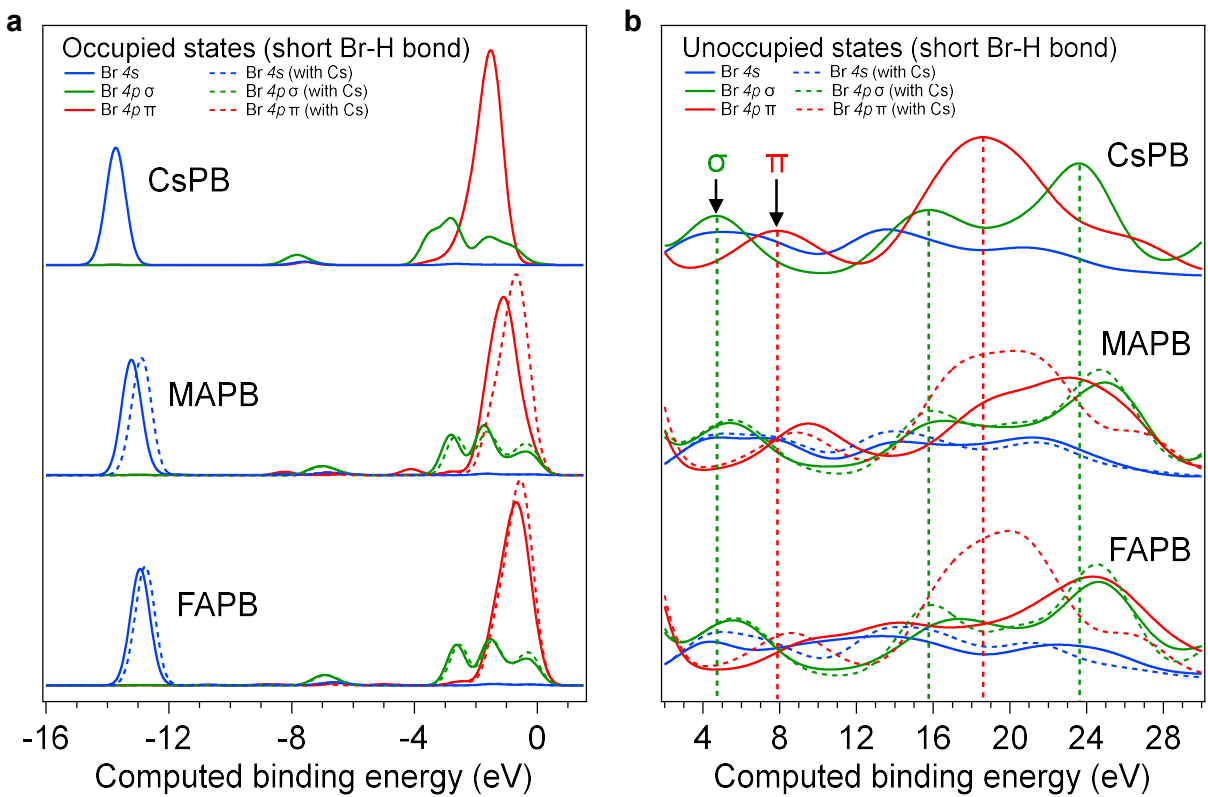

**Figure 3 | Calculated ground-state bromine *4s*, *4p* σ and *4p* π projected density of states of CsPB, FAPB and MAPB**. **a**, Occupied states are enumerated with negative binding energies and are Gaussian-broadened by 0.3 eV. The top of the valence band is aligned to 0 eV binding energy. **b**, Unoccupied states are Gaussian-broadened by 1.0 eV to emphasize the main features of interest. The two π components have been averaged together to yield one Br *4p* π distribution. Projected DOS corresponding to strongly hydrogen-bonded bromine with organic molecules are shown. Bromine PDOS associated with cesium-substituted MAPB/FAPB are shown with dashed lines. The vertical dashed lines are guides to the eye and mark CsPB features of interest which can be compared to similar features for original and cesium-substituted MAPB and FAPB.

To investigate the physicochemical interactions underlying the σ-π offset trend, MA$^+$/FA$^+$ were artificially substituted with Cs$^+$ in the geometries of MAPB/FAPB while maintaining the Br-Pb sublattice structure. Remarkably, when the organic A-cations are substituted by Cs$^+$, the sharpness and energetic positions of both the π and σ contributions in the unoccupied states become CsPB-like, even though the structural models have not been relaxed. The σ-π offset for CsPB is of comparable magnitude to the offsets for Cs-substituted MAPB/FAPB, and no systematic trend is observed for short/long H-Br bond distances (ST 1). Hence, these artificial investigations provide computational evidence that H-bonding in MAPB/FAPB has a strong influence on the conduction band and an electronic coupling exists between the A-cations and the Br-Pb sublattice. This is consistent with reported electronic coupling in the unoccupied states of MAPI[23]. Furthermore, the influence on the σ and π symmetry contributions of the Br PDOS indicates that the magnitude of the σ-π offset in the XAS spectra is influenced by the strength of H-bonding interaction.

We turn our attention to possible manifestation(s) of A-cation influence in the occupied states and in the crystal structure. Our calculations suggest the occupied Br $4p$ PDOS near the valence band maximum (VBM) is weakly affected by A-cation type (Fig. 3a). To investigate further, we analyze the Br $K$ valence-to-core (VtC) XES spectra (Fig. 4 right inset) since the Br $K\beta_2$ transition, shown in the RXES map (Fig. 1a) and measured simultaneously with the $K\beta_1$ transition which yields the HERFD-XAS spectrum, carries information on the Br $4p$ states in the valence band. We fit a Voigt peak to the main VtC line and observe that the ~13462 to ~13472 eV region is similar for all compounds, in agreement with calculation. This finding, combined with our earlier finding of mostly Br-Pb σ-states near the CBM (Fig. 2c-e), could explain why band-edge carrier dynamics in APB crystals are independent of the A-cation type[24]. Additionally, we observe that the relative VtC intensity of FAPB is higher than MAPB/CsPB (Fig. 4 center inset), given the same number of bromide X-ray emitters (normalized to $K\beta_{1,3}$). This shows the number of electrons in Br $4p$ is higher for FAPB and suggests the Br-Pb bond for FAPB is more ionic. We confirm this observation by examining the chemical shift, finding an increasing Br-Pb ionicity trend (CsPB→MAPB→FAPB) which matches the σ-π offset trend. Technical details are found in SN 3. In principle, bond ionicity originates from the electronegativity difference between lead and bromine, hence the trend is unanticipated and reveals the A-cation influence on carrier mobility, which has previously been exponentially correlated with halide-lead bond ionicity[25]. Additionally, the $K\beta_1$ trend matches the XAS onset trend of FAPB→MAPB→CsPB, thus we see that there is a correlation between the strength of A-cation coupling and Br-Pb bond ionicity, which determines relative shifts in all energy levels referenced to Br *1s* and has implications for energy level matching at device interfaces.

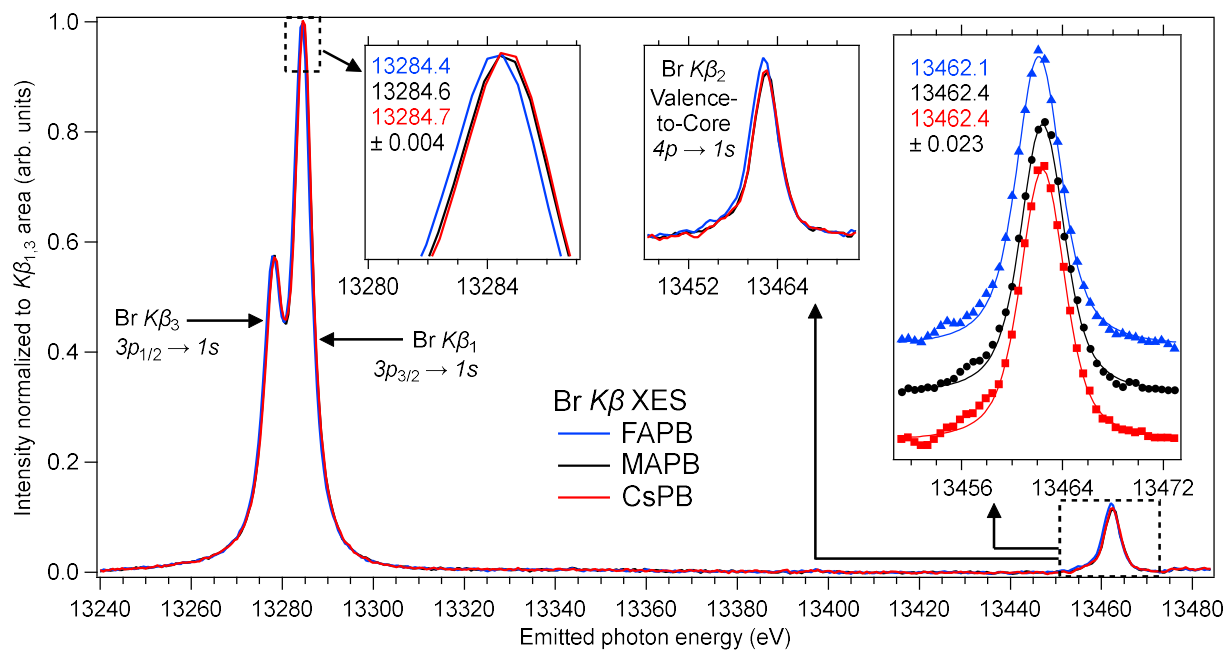

**Figure 4 | X-ray emission spectra of FAPB, MAPB and CsPB.** The $K\beta_1$ and $K\beta_2$/VtC peaks are fitted with Voigt peaks; the fitted values and uncertainties are displayed in the insets. The Voigt peak fit of the main VtC emission line is intended to extract the energy position of the peak maximum. The center inset shows VtC intensities associated with normalized $K\beta_1$ intensities.

We utilize hard X-ray PES (HAXPES) measurements for a complementary investigation of the VBM DOS, where differences in spectral profiles could be assigned to differences in lead contributions as we have shown the Br $4p$ PDOS of FAPB, MAPB and CsPB are similar. We observe that the A-cation influences the DOS near the VBM (SF 4) and deduce that the A-cation type modifies the degree of Pb $6s$/$6p$ hybridization with the Br $4p$ states, which may be relevant for visible light absorption/emission and Rashba band-splitting effects. Further analysis and discussion are found in SN 4.

Through a combined analysis of Pb $L_3$ RXES, AIMD-derived bond angles/distances (SF 5) and Goldschmidt tolerance factor (GTF), we examine A-cation effects in the crystal structure[26]. Our X-ray diffraction measurements confirm the crystal structures are as expected (ST 3) and Pb $L_3$ RXES confirms the unit cell symmetry increases with increasing GTF. We find that cooperative PbBr$_6$ octahedral tilting shows the largest geometrical variation between the APB compounds, but were unable to relate that mechanistically to the σ-π trend which instead is related directly to hydrogen bonding. We derive relative GTF ratios for MAPB : FAPB (0.98) and CsPB : FAPB (0.79) and find they nearly match the relative main-edge width MAPB : FAPB ratios of 0.96 (measured) and 0.95 (calculated) and CsPB : FAPB ratios of 0.83 (measured) and 0.83 (calculated). This is a positive and potentially linear correlation between relative GTF and relative conduction band width which applies to the APB compounds and possibly all HaP compounds. Consequently, we deduce that the GTF can be regarded as both a measure of H-bonding strength in MA$^+$/FA$^+$ (and its absence in Cs$^+$) as well as an approximate measure of octahedral tilting. Technical details can be found in SN 5. We summarize the numerical quantities obtained through our experimental and computational work and aforementioned correlations in Fig. 5.

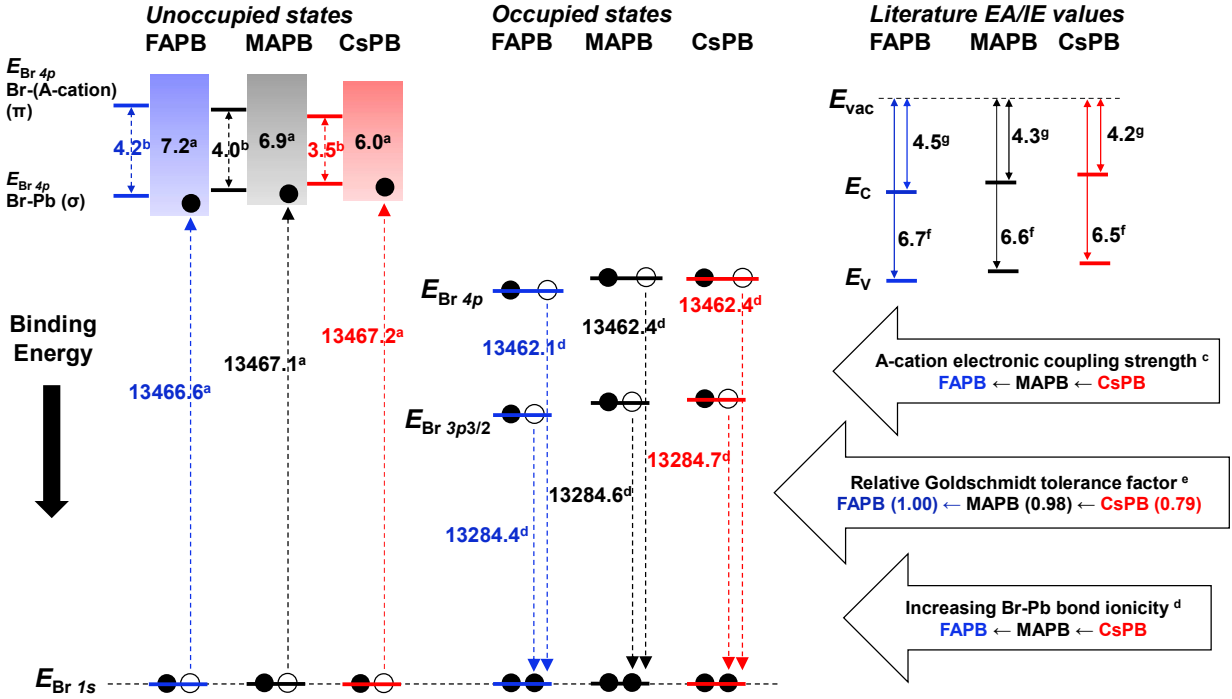

**Figure 5 | Summary of energy level positions derived from experiment and calculation, and correlations found.** One-electron transitions are shown for the X-ray absorption and emission processes. The origin of each numerical value or correlation is indicated with a superscript identifier: (a) Br $K$ HERFD-XAS (experiment), (b) Br $K$ XAS (calculation), (c) Br PDOS (calculation), (d) Br $K$ XES (experiment), (e) Goldschmidt tolerance factor (literature), (f) UPS (literature) and (g) inverse PES (literature). A-cation electronic coupling strength refers to the strength of coupling between the A-cation and Br-Pb sublattice. Electron affinity and ionization energy are abbreviated as EA and IE, respectively, and their values from the literature are shown, for comparative purposes[11]. The connection between this work and the work reported in the literature is described in the text.

Using Fig. 5, we briefly discuss the implications of our findings for energy level matching at device interfaces and slow hot electron cooling. We note that direct and inverse PES measurements of device-relevant energy level positions show a trend which systematically matches our Br $K$ XES and HERFD-XAS trends[11]. The interface energetics are sensitive to A-cation coupling strength; even a ~100 meV (~4 $k_B$T at room temperature) shift in the energy levels will substantially affect interfacial carrier transport mechanisms where current density depends exponentially on the magnitude of the energy barrier/offset (i.e. thermionic emission). State-of-the-art HaP solar cells and light-emitting diodes typically feature a mixture of MA$^+$, FA$^+$ and Cs$^+$ A-cations and depth-profiling with HAXPES has revealed spatial variations of A-cations near the surface[27,28]. Our study reveals that the A-cation(s) present, intended or not, in the few perovskite layers adjacent to the interface are expected to influence the interface energetics and hence device performance. Slow hot carrier cooling in HaPs is of interest due to the technological potential of hot-carrier solar cells[29]. Time-resolved PL studies have yielded evidence for at least two distributions of states, consistent with our finding of the Br *4p* σ and π PDOS in the conduction band, where the states which are excited at higher optical energies show faster PL decay[12]. This observation suggests the π states exhibit a higher carrier thermalization rate. Furthermore, we observe a positive correlation between hot carrier decay time constant and σ-π offset and hypothesize that a larger σ-π offset moves the π states further away from the CBM and reduces the average carrier thermalization rate. Technical details are presented in SN 6. The potential connection between the σ-π character of the conduction band and polaron-like carrier transport is discussed in SN 7.

In conclusion, through a joint experimental-computational investigation, we have shown that the energetic width of the conduction band depends on the strength of electronic coupling between the A-cation and the lead-bromide sublattice. Our finding of the σ-π offset in the conduction band provides an additional mechanism to consider, in the search for the origins of slow hot carrier cooling and polaronic transport. In both the occupied and unoccupied states, we find a correlation between higher coupling strength and higher bromide-lead bond ionicity and a decreased energy offset between a given energy level and Br *1s*, and discuss the implications for energy level matching at perovskite device interfaces. We foresee the spectroscopic approach we have employed in this work being utilized to study A-cation effects in other halide perovskites and potentially non-halide perovskites as well, thus unravelling unexplored structure-property relationships in these materials and furthering the development of optoelectronic devices.

## Acknowledgements

The RXES measurements were carried out at beamline P64 of the synchrotron facility PETRA III at DESY, a member of the Helmholtz Association (HGF).  The von Hamos-type hard X-ray spectrometer was realized in the frame of projects FKZ 05K13UK1 and FKZ 05K14PP1.  We thank DESY for the provision of beamtime (Proposal No. I-20181028 EC, I-20190356 EC).  Experiments at PETRA III have been supported by the project CALIPSOplus under the Grant Agreement 730872 from the EU Framework Programme for Research and Innovation HORIZON 2020.  The HAXPES measurements were performed at the HIKE end-station attached to the beamline KMC-1 of the synchrotron facility BESSY II.  We thank HZB for the provision of beamtime (Proposal No. 192-08827).

SMB and GJM thank the Swedish Research Council (contract 2018-05525) for financial support.

HR, GJM, DP and SM acknowledge the Swedish Research Council (grant # 2018-06465 and # 2018-04330) and the Swedish Energy Agency (grant # P43549-1) for funding.

PKN and JA acknowledge support from the Department of Atomic Energy, Government of India, under Project Identification no. RTI 4007 and Science and Engineering Research Board India core research grant (CRG/2020/003877).

MO and KC acknowledge support from the European Union's Horizon 2020 Research and Innovation programme under the Marie Skłodowska-Curie grant agreement No 860553, and the Swedish energy agency (contract 2017-006797).  The calculations were enabled by resources provided by the Swedish National Infrastructure for Computing (SNIC) at the Swedish National Supercomputer Center (NSC), the High Performance Computer Center North (HPC2N), and Chalmers Centre for Computational Science and Engineering (C3SE) partially funded by the Swedish Research Council through grant agreement no. 2018-05973.

We thank Sigurd Wagner (Princeton) for discussions related to the mechanism of hot carrier cooling in HaPs.

## Author contributions

G.J.M. and S.M.B. conceived and designed the experimental study.  P.K.N. grew the single crystals. G.J.M., S.M.B. and A.K. performed the RXES measurements and analyzed the data.  D.P., S.M. and G.J.M. performed the HAXPES measurements; D.P. and G.J.M. analyzed the data.  J.A. performed the XRD characterization and analysis, under the supervision of P.K.N.  C.K. performed AIMD simulations, PDOS and XAS calculations for the snapshots and analyzed all computational results.  G.J.M., S.M.B., C.K. and M.O. analyzed the XAS.  G.J.M. proposed the links between the findings and questions of technological relevance.  G.J.M. and S.M.B. wrote the first draft of the manuscript.  All authors contributed to revisions of the manuscript.  G.J.M., H.R., M.O. and S.M.B. supervised the project.

## Competing interests

The authors declare no competing interests.

**Methods**

**APB crystal growth.**  The MAPB and FAPB crystals were solution-grown using methods reported by us previously[30,31].  We grew the CsPB crystals using a method reported by others[32].

**X-ray diffraction.**  X-ray diffraction measurements were performed on FAPB and CsPB single crystals at room temperature using a Rigaku diffractometer with graphite-monochromated molybdenum K$\alpha$ radiation ($\lambda$= 0.71073 Å).  Data processing was performed with the CrysAlisPro software.  Empirical absorption correction was applied to the collected reflections with SCALE3 ABSPACK and integrated with CrysAlisPro. The structures of the single crystals were solved by direct methods using the SHELXT program, and refined with the full-matrix least-squares method based on $F^2$ by using the SHELXL program through the Olex interface.  The crystallographic parameters for MAPB are taken from reference[30], since a similar growth procedure was used by the same preparer.

**Hard X-ray absorption and emission spectroscopy.**  The measurements were performed (on APB crystals from the same batch characterized with XRD) at beamline P64 of the synchrotron facility PETRA III[33].  Incident photon energy calibration at the Pb $L_3$ edge was performed with metallic lead foil.  The photon flux incident on the sample at ~13 keV was ~3 x $10^{11}$ photons s$^{-1}$.  The X-ray spot size, measured with the X-ray eye, is ~220 μm x 100 μm.  Total FY-XAS measurements were recorded with a passivated implanted planar silicon (PIPS) detector (Canberra).  Resonant XES maps were recorded using a von Hamos-type hard X-ray crystal spectrometer mounted in a Bragg scattering configuration.  The spectrometer featured 8 crystal analyzers; the third order reflection of the Si(220) crystal analyzers was used for Pb $L\alpha_{1,2}$ measurements and the fourth order reflection was used for Br $K\beta_{1,2,3}$ measurements[20].  Energy calibration of the two-dimensional detector images was performed with custom Python-based software written by Dr. Aleksandr Kalinko.  Bromine $K$ (~13.47 keV) and lead $L_3$ (~13.04 keV) HERFD-XAS spectra were generated as averaged intensities of the slice cut from RXES maps.  The energetic width of the slice on the emitted energy axis (HERFD linewidth) was 1.6 eV and the cut was done through the RXES maximum.  The HERFD linewidth was selected based on a balance between energy resolution and signal-to-noise (Supplementary Figure 6).  The off-resonant, higher incident energy half of the RXES map was used to generate the X-ray emission spectrum.  The energy resolution of XES is determined by the ~0.3 eV photon bandwidth of the Si(311) monochromator and the ~1.3 eV FWHM Gaussian broadening of the spectrometer.  All measurements on the APB single crystals were performed inside a Linkam T95 heating/cooling stage at room temperature.  The Linkam was initially purged with nitrogen; the crystals remained in a nitrogen atmosphere during the experiments.  Halide perovskite compounds are known to degrade upon irradiation with energetic beams[34,35].  We optimized the beamline, spectrometer and measurement settings to obtain at least two iterations of HERFD-XAS spectra, recorded sequentially and from the same (initially fresh) spot of a single crystal sample, which do not show observable differences due to beam damage.  The measurement duration of each Br $K$ and Pb $L_3$ spectrometer iteration/run was optimized to be ~12 minutes long.  Supplementary Figure 7 shows Br $K$ and Pb $L_3$ HERFD-XAS spectra, recorded from single crystal MAPB, associated with pristine (iterations 1 and 2) and degraded MAPB (long beam exposure, iteration 13).

**Hard X-ray photoelectron spectroscopy.**  Valence band and core level PES measurements were performed at the HIKE endstation attached to the beamline KMC-1 of the synchrotron facility BESSY II [36]. With an excitation energy of 4000 eV, the measurements are expected to be more bulk-sensitive and as-grown (un-cleaved) crystals were used.  The overall energy resolution of PES, determined from the FWHM of Gaussian-broadened Au $4f_{7/2}$ photoelectron lines, is 0.3 eV and is governed primarily by the photon bandwidth of the Si(311) monochromator.  Binding energy calibration was performed by recording Au $4f$ spectra from a reference gold foil and setting the binding energy of $E_{\text{Au}4f7/2}$ to 84.0 eV. Multiple iterations of core level spectra were recorded, to distinguish the onset of beam damage.  See Supplementary Note 8 and Supplementary Figures 8 and 9 for further details.

**Calculations.**   We performed periodic DFT calculations using the Quickstep code in the CP2K package[37–39]. We used the Perdew-Burke-Ernzerhof (PBE) generalized gradient approximation as the exchange-correlation (XC) functional[40].  In addition, we have included a Van der Waals correction using Grimme's D3 method[41].  The geometry optimizations were performed using the Gaussian plane-wave (GPW) method with Goedecker-Teter-Hutter (GTH) pseudopotentials as well as TZVP-MOLOPT-GTH (for carbon, nitrogen, hydrogen) and DZVP-MOLOPT-GTH (for lead and bromine) basis sets[42–44].  An energy cut-off of 600 Ry was used along with a multi-grid consisting of five grids. The starting geometries for MAPB, FAPB and CsPB are (cubic, 3x3x3), (cubic, 3x3x3) and (orthorhombic, 2x2x4), respectively, taken from the literature[45–47].  All electronic structure calculations were performed at the gamma point of the super cells.  Ab initio molecular dynamics simulations were carried out with the same computational parameters within an NPT ensemble at 300 K and 0 atm.  The MD simulations were performed for a minimum of 30 ps using a 0.5 fs time-step.

X-ray absorption spectra were calculated using the Gaussian augmented plane-wave (GAPW) implementation of core-level spectroscopy in CP2K, which allows for a mixed pseudopotential and all-electron description [22,48]. The bromine, hydrogen, carbon and nitrogen atoms were treated at an all-electron level and described with 6-311Gxx basis sets while the GTH pseudopotential with TZVP basis sets were used for the lead and cesium atoms.  To simulate core-hole relaxation effects, we use the transition potential method with a half-core hole (TPHH) at the Br *1s* state.  Separate spectral simulations were carried out for each bromine atom in the supercell, where the supercells are obtained from geometry optimization as well as AIMD snapshots (three or four configurations are selected for FAPB, MAPB or CsPB with 10 ps time differences in the AIMD trajectory and about 100 Br atoms for each configuration).  Individual spectra are then averaged together to generate aggregate spectra.  The calculated, discrete XAS spectra are convoluted with a Gaussian function with a broadening parameter or σ of 1.0 eV (corresponding to FWHM of 2.355 eV) to simulate experimental spectra.  Absorption energies obtained from simulation are underestimated in comparison to experimental data; this is a known limitation of the TPHH approximation.  Consequently, an ad-hoc rigid shift of +194.3 eV (~1.5%) was added to all calculated Br $K$ XAS spectra for direct comparison to experiment.